 \documentstyle[twocolumn,prl,aps]{revtex}

\begin{document}
\draft
\preprint{}

\twocolumn[\hsize\textwidth\columnwidth\hsize\csname@twocolumnfalse%
\endcsname

\title{Frustration induced Raman scattering in CuGeO$_3$}
\author{V. N. Muthukumar$^1$, Claudius Gros$^1$, 
        Wolfgang Wenzel$^1$, Roser Valent\'\i$^1$,\\
        P. Lemmens$^2$, B. Eisener$^2$, G. G\"untherodt$^2$,
        M. Weiden$^3$, C. Geibel$^3$, F. Steglich$^3$
       }
\address{$^1$ Institut f\"ur Physik, Universit\"at Dortmund,
         44221 Dortmund, Germany\\
        }
\address{$^2$ 2. Physikalisches Institut, RWTH Aachen,
         52056 Aachen, Germany
        }
\address{$^3$ FB Technische Physik, TH-Darmstadt, 
         Hochschulstr. 8, 64289 Darmstadt
        }
\date{\today}
\maketitle
\begin{abstract}
We present experimental data 
for the Raman intensity in 
the spin-Peierls compound CuGeO$_3$ 
and theoretical calculations from
a one-dimensional frustrated spin model.
The theory is based on
(a) exact diagonalization and (b) a 
recently developed solitonic mean field theory. 
We find good agreement between the 1D-theory 
in the homogeneous phase and evidence 
for a novel dimerization of the Raman operator 
in the spin-Peierls state.
Finally we present evidence for a
coupling between the interchain exchange, 
the spin-Peierls order parameter and 
the magnetic excitations along the chains.

\end{abstract}
\pacs{42.65.-k,78.20.-e,78.20.Ls}
]  
Low-dimensional spin systems exhibit many
unusual properties resulting from
quantum and dimensionality effects.
An example is the continuum of spin-wave
excitations in quantum one-dimensional
(1D) spin systems which has been predicted
for a long time \cite{Mueller} and has
recently been confirmed by neutron scattering
experiments \cite{KCuF3} on KCuF$_3$.

Quantum 1D spin systems also show a variety
of instabilities. Of particular interest is the
spin-Peierls (SP) phase due to residual
magnetoelastic couplings \cite{Beni_Pincus},
which leads to the opening of a gap in the
spin excitation spectrum.
The discovery \cite{Hase} of the spin-Peierls 
transition below $T_{SP}=14$ K 
in an inorganic compound, CuGeO$_3$, 
has attracted widespread attention. This compound
consists of chains 
of spin-1/2 Cu$^{2+}$ ions coupled by 
antiferromagnetic superexchange via 
the oxygen orbitals \cite{Hase,Nishi}. 
The Cu ions lie along
the crystallographic c-axis and
the exchange along the chains can
be modeled by the 1D Hamiltonian
\begin{equation}
H = J\sum_i [ (1+\delta(-1)^i){\bf S}_i\cdot{\bf S}_{i+1}
            +    \alpha       {\bf S}_i\cdot{\bf S}_{i+2}
            ]~,
\label{H}
\end{equation}
where $\delta$ is the dimerization parameter that vanishes above
$T_{SP}$ \cite{Castilla,Riera}.
The special geometry \cite{Castilla,Braden}
of the super\-exchange
path in CuGeO$_3$ leads to a small value of the
exchange integral $J\approx150$K and a substantial
n.n.n. frustration term $\sim\alpha$ which competes
with the n.n. anti\-ferromagnetic exchange.
The interchain couplings have been
estimated to be small, $J_b\approx0.1J$ and
$J_a\approx-0.01J$ for the interchain exchange
constants along a- and b- directions, respectively
\cite{Nishi}.

The phase diagram of $H$ in Eq.\ (\ref{H})
has been calculated using the 
density-matrix renormalization-group method \cite{Chitra}.
For $\delta=0$ and $\alpha<\alpha_c\approx0.2411$,
the ground state is gapless and renormalizes to the
Heisenberg fixed point.  For $\alpha=0.5$
and $\delta=0$, the ground state is given by a valence-bond
state and a gap of order $J/2$ induced by frustration is
present. 
While the evaluation of the dynamical properties of (\ref{H})
is a challenge to theory, the rich phase diagram can be explored
by a variety of interesting experiments. In this context, the
substantial value of the n.n.n. exchange integral
in CuGeO$_3$ allows the experimental investigation
of the effects of competing interactions in
a low dimensional magnet, both in the uniform and in the
spin-Peierls state.  

An experimental method particularly suited for the
study of magnetic excitations in an
antiferromagnet is two magnon
Raman scattering.
For CuGeO$_3$, the Raman operator in $A_{1g}$ 
symmetry \cite{Fleury}is proportional to 
\begin{equation}
H_R = \sum_i\,(1+\gamma(-1)^i)\,{\bf S}_i\cdot{\bf S}_{i+1}~.
\label{H_R}
\end{equation}
In the homogeneous state ($\delta=\gamma=0$) 
the interaction Hamiltonian 
commutes with the Heisenberg Hamiltonian 
for the case $\alpha=0$ and 
there would be no Raman scattering.
However when $\alpha \neq 0$, the model (\ref{H}) leads to magnetic
Raman scattering due to the presence of competing interactions
which can be observed experimentally.
Previous fits to experiments \cite{Raman} 
based on (\ref{H})  assumed $\alpha=0$.  Next, we note the presence 
of the factor $\gamma$ in Eq.\ (\ref{H_R}) which
appears because the exchange integral is sensitive 
to the inter-ionic distance.  

Both features mentioned above are taken into
consideration in this Letter where we
present experimental data as well as the first
theoretical results for the Raman intensity 
in one-dimensional spin systems obtained 
from (a) exact diagonalization
studies of chains with $N_s \leq $28 sites and
(b) a newly developed solitonic mean field
theory. Our results for $T>T_{SP}$ are obtained with $\delta = 0$ in
Eq.\ (\ref{H}), where the mean field theory provides an 
analytical framework to
obtain expressions for the Raman intensity at finite
temperatures. Below $T_{SP}$,
we present first experimental evidence for a qualitative change 
in the Raman operator at the SP transition
through the appearance of the factor $\gamma$ in Eq.\ (\ref{H_R}). 
Here, we do not attempt to calculate
$\gamma$ microscopically, but by comparing
with experiment. We
find that $\gamma\approx0.12$ and this value of
$\gamma$ affects the shape of the Raman spectrum in the 
SP phase dramatically. Indeed, we find that the
experimental results for $T<T_{SP}$ cannot be
explained by Eq.\ (\ref{H}) assuming $\gamma$ to be zero.
A detailed comparison of 
experiment and theory, both below and above
the SP transition, is presented. 


We first give a brief account of 
the theoretical and experimental methods 
before discussing the experimental data. 
The Raman intensity at zero temperature is given by
\begin{equation}
I_R(\omega) \,=\, -{2\over\pi}Im\,
\langle0|H_R{1\over \omega+i\epsilon - (H-E_0)}H_R|0\rangle,
\label{I_R}
\end{equation}
where $E_0$ is the ground-state energy, $H$ the
Hamiltonian given by Eq.\ (\ref{H}) and $\epsilon\rightarrow0+$.
For a numerical evaluation of Eq.\ (\ref{I_R}), we scale the
Hamiltonian $H=cX+d$ such that the eigenvalues of
the rescaled Hamiltonian $X$ are in the interval
$[-1,1]$. We define a rescaled energy and frequency
by $E_0=cx_0+d$ and $\omega=cx+d$ and expand 
$I_R(x)$ in terms of Tschebycheff polynomials, 
$T_l(x)$:
$I_R(x) = (1-x^2)^{-1/2}\sum_{l=0}^{N_p} a_l T_l(x+x_0)$
with $a_l=(2-\delta_{l,0})/\pi\langle0|H_RT_l(X)H_R|0\rangle$
and $N_p\rightarrow\infty$. The quantities $a_l$ are evaluated
recursively.  This procedure, the {\sl kernel
polynomial approximation}, is an established
\cite{Silver} and numerically stable method for the
evaluation of the density of states \cite{note_2}. For large
values of $N_p$, the resulting spectral weight consists of
a series of very sharp peaks that become delta-functions
in the limit $N_p\rightarrow\infty$. Generally we find
$N_p=100$ convenient for comparison with experiment.

We also calculate the Raman intensity above $T_{SP}$ using
a solitonic mean field theory. This method, originally
suggested for the n.n. Heisenberg chain \cite{mean_field}, 
is based on a transformation of the Heisenberg Hamiltonian 
to a Hamiltonian describing
the dynamics of antiferromagnetic domain walls 
(solitons or spinon excitations) 
and a subsequent Jordan-Wigner transformation. 
The method reproduces the exact solutions at both the 
Ising and XY-limits. For the Heisenberg
model it leads to correct asymptotic behavior of 
dynamical correlation functions. 
The solitonic mean field theory leads to a
ground state described by the Hamiltonian 
$H_S = \sum_k E_k c^\dagger_k c_k$, 
where the $c_k$'s are quasiparticle (spinon) operators that are
linear combinations of the soliton operators. 
The mean field dispersion relation
$E_k = (1+{2 \over \pi}) J |\cos ka|$ compares very well
with the des Cloizeaux-Pearson spectrum
${\pi \over 2} J |\cos ka|$ obtained from Bethe ansatz
\cite{bethe}. We generalize this method to the case of 
the n.n.n. chain. 

The absence of Raman scattering in the n.n. chain is
understood as $H_S$ conserving the number of spinons. This picture
changes when $\alpha \neq 0$. We find that the inclusion of the n.n.n.
term generates the following processes: 
(i) two-spinon scattering terms
that lead to a renormalization of the spinon velocity and
(ii) four-spinon creation terms that generate two-magnon 
Raman scattering in the n.n.n. chain.
A perturbative treatment of the four-spinon terms allows us
to obtain expressions for the Raman intensity $I_R(\omega)$ 
in the homogeneous phase at all temperatures.

%

The experiments were performed
using the excitation line $\lambda=514.5$-nm 
of an Ar-laser with a laser power of 2.7mW. 
We ensured that the incident radiation does not
increase the temperature of the sample by more than 1.5 K.
We used a DILOR-XY spectrometer and a nitrogen cooled CCD 
(back illuminated) as a detector in a quasi backscattering 
geometry with the polarization of incident and scattered light parallel 
to the c-axis and the Cu-O chains, respectively. 
Details of the experiment will be published elsewhere \cite{Lemmens_long}.


In Fig.\ \ref{T20}, we present the data for the 
two-magnon Raman continuum in the homogeneous 
state at $T=20$ K. 
Phonon lines \cite{Raman} at 184 cm$^{-1}$ 
and at 330 cm$^{-1}$ are
subtracted from the experimental 
data (squares). The shoulder observed at 
$\sim390{\rm cm}^{-1}$ is presently not understood.
We interpret the two-magnon continuum 
as scattering intensity caused by the creation of
four spinon excitations. The Raman intensity from such 
a scattering process calculated from our solitonic 
mean field theory is shown in Fig.\ \ref{T20}
(dashed-dotted line), where we use  
$J=150$ K $\sim$ 104 cm$^{-1}$
and $\alpha=0.24$\cite{Castilla}. 
The maximum theoretical value
of the Raman intensity is normalized to the 
experimental value from which a Rayleigh tail
and a uniform background 
of 50 counts are subtracted.
The maximum of the experimental data is 
situated at slightly larger energies,
which could be a consequence 
of a slight misfit of the parameters 
used.  For instance, we find that $J=160$ K and 
$\alpha=0.2$ (dotted line)
improves the agreement between theory and experiment. 
The data presented in Fig.\ \ref{T20}
provide substantial evidence that
a 1D Hamiltonian of type (\ref{H}) can indeed
account for the observed Raman continuum owing to the presence 
of the n.n.n. frustration term. This continuum should also be seen
in neutron scattering experiments above $T_{SP}$.


In Fig.\ \ref{theory} we present the numerical results
for the dimerized state, 
$\delta=0.03$ \cite{Castilla}, and a
finite-size analysis (in the inset).
The numerical result for $\gamma=0$ 
(upper curve, dashed line) is surprisingly flat 
and can be approximated (upper curve, solid line)
by the expression
\begin{equation}
I_R(\gamma=0,\omega) = A\theta(\omega-2\Delta)
\left(1-\tanh[2(\omega-\omega_0)]\right),
\label{I_0}
\end{equation}
with the values of the parameters
$2\Delta\approx30{\rm cm}^{-1}$,
$\omega_0\approx312{\rm cm}^{-1}$ being
determined by a fit to the numerical data
(A is a normalization constant). 
Below T$_{SP}$, a gap
opens up and the DOS has a singularity at the
lower edge \cite{Raman,Lemmens}. Therefore, one
expects a peak arising from this singularity.
However, this singularity is removed by
matrix element effects arising from the Raman
operator \cite{note_xy}.

The $\gamma$-dependence of the Raman intensity
is strong, as can be seen in
Fig.\ \ref{theory}, where we plot the
data for $\gamma=0.12$ (lower curves)
In order to understand this large matrix-element
effect we have examined in detail 
(for $N_p=1000$) the relative weight 
$\rho(\gamma,E_i)$ of the
individual poles (at energies $E_i$)
contributing to the Raman intensity
for systems of size $N_s=20,\ 24$ and 28.
We have found that the basis of the observed 
matrix-element effect lies in the remarkable fact 
that for each pole there is a certain
$\gamma_0(E_i)$ at which the intensity
of the pole actually vanishes, i.e.
\begin{equation}
\rho(\gamma,E_i) = {I_{\gamma=0}\over I_\gamma}
\left({\gamma-\gamma_0(E_i)\over\gamma_0(E_i)}\right)^2,
\label{erasor}
\end{equation}
where $I_\gamma$ is a normalization constant defined
by $\int_0^{\omega_c}d\omega\rho(\gamma,\omega)=1$
(we choose $\omega_c=6J$).
From the numerical data we found that 
$\gamma_0(E_i)$ can be approximated by
$\gamma_0(E_i)\approx\delta+\kappa E_i$, 
where $\kappa\approx4/3000{\rm cm}$.
Combining (\ref{I_0}) and (\ref{erasor}),
we obtain an analytic approximation for the Raman
intensity in the dimerized state,
\begin{equation}
I_R(\gamma,\omega)\approx\rho(\gamma,\omega)I_R(0,\omega)
\label{analytic}
\end{equation}
This formula reproduces the
numerical results well (see Fig.\ \ref{theory}).


In Fig.\ \ref{T5}, we present data
for the spin-Peierls phase at $T=5$K.
Let us first discuss the line at 30 cm$^{-1}$. We
find that the analytic curve for $\gamma=0.12$
reproduces this peak well (solid line in
Fig.\ \ref{T5}). Note that 
here $\gamma$ is the only free parameter in the
theory. Choosing $\gamma=0$ instead 
would result in a complete disagreement with 
experiment (compare Fig.\ \ref{theory}). 
The line at 30 cm$^{-1}$ is known
to be of 1D magnetic origin
(the value 30  cm$^{-1}$ is indeed twice 
the one-magnon gap obtained from neutron scattering 
\cite{Nishi}). This suggests that this peak
arises from the spin-Peierls Hamiltonian (\ref{H})
in combination with matrix-element effects
of the Raman-operator (\ref{H_R}).
Therefore we conclude that Fig.\ \ref{T5}
provides strong evidence for the
dimerization of $H_R$ below $T_{SP}$
leading to substantial matrix-element effects.

%

Let us now consider the two lines at 226 cm$^{-1}$ 
and at 104 cm$^{-1}$ observed below $T_{SP}$
(see Fig.\ \ref{T5}). The assignment of the 
104 cm$^{-1}$ line is still controversial
and will not be discussed here.
The 226 cm$^{-1}$ line has been assigned previously
to be of magnetic origin \cite{Raman}. A
classical (non-interacting) spin-wave calculation, 
using the measured magnon dispersion in c-direction
\cite{Nishi} produces a peak at 226 cm$^{-1}$
\cite{Lemmens}. However, it is well 
known that in ideal one-dimensional systems,
magnons do not behave classically but 
form a continuum of excitations\cite{Mueller}.
Consequently one would not expect
a sharp peak, like the one observed 
at 226 cm$^{-1}$ to occur in a one-dimensional system.
This expectation is borne out by our numerical
data (solid line in Fig.\ \ref{T5}), which
does not show the 226 cm$^{-1}$ peak.
It is clear that to explain the presence of this peak, 
one has to go beyond a one-dimensional spin model.

The magnitude of the interchain coupling $J_b$ 
in CuGeO$_3$ is comparable to the spin-Peierls 
temperature (14 K). One might therefore expect the 
chains to become correlated for temperatures
around $T_{SP}\approx J_b$, resulting in the
appearance of a well defined magnon branch, which
leads to the observed peak at 226 cm$^{-1}$.
In Fig.\ \ref{rel},
we present the temperature dependence
of the intensity of the 226 cm$^{-1}$
line (for comparison we include
in Fig.\ \ref{rel}, two other Peierls-
active Raman lines). One clearly sees
that this line becomes active only
below $T_{SP}$. Thus we conclude
that the interchain coupling becomes
relevant only below $T_{SP}$. This
conclusion is in agreement with
recently presented neutron scattering
data \cite{Kakurai}, which shows a
pronounced change in the spectrum
below $T_{SP}$.


In conclusion, we find good agreement between
the experimental magnetic Raman spectrum 
in the homogeneous phase and
the one-dimensional frustrated spin model for
CuGeO$_3$. The importance of the competing
interactions for the occurrence of magnetic
Raman intensity in the homogeneous
phase has been pointed out and is
consistent with the absence of
inelastic Raman scattering in
1D spin compounds without frustration such as
KCuF$_3$ \cite{Yamada}.
For the spin-Peierls state we find that
observation of a sharp line at the
spin-triplet excitation energy of 30 cm$^{-1}$
indicates a dimerization of the Raman operator.
Finally, the observed Peierls-active line at 226 cm$^{-1}$ 
indicates the appearance of a well defined magnon branch 
below $T_{SP}$. 


\acknowledgements
This work was supported through the Deutsche 
For\-schungs\-gemein\-schaft, 
the Graduierten\-kolleg ``Fest\-k\"{o}rper\-spekt\-roskopie'',
SFB 341 and SFB 252,
and by the BMBF 13N6586/8,

%
%
%
\begin{figure}
\caption{Experimental (squares) and theoretical
         results (lines) for the Raman intensity
         in the homogeneous phase of CuGeO$_3$ at 
         $T=20$ K. Shown are the results from
         the solitonic mean field theory
         at T = 20 K ($\delta=0=\gamma$) 
         and for two sets of parameters
         (a) $J=104$ cm$^{-1}$,
         $\alpha=0.24$ \protect\cite{Castilla} 
         (dashed-dotted line), and
         (b) $J = 119$ cm$^{-1}$, $\alpha =0.2$ (dotted line).
\label{T20}
             }
\end{figure}
\begin{figure}
\caption{The numerical results for $N_s=28$ (dashed lines),
         $\delta=0.03$, $\alpha=0.24$, $N_p=100$ 
         and $\gamma=0$ (upper curve) and $\gamma=0.12$
         (lower curve). The numerical results are compared
         with the analytic formula (\protect\ref{analytic})
         (solid lines).  Inset: A comparison of numerical 
         results for $N_s=28$ (dashed line) and $N_s=24$ (dotted
         line) and $\gamma=0$.
\label{theory}
             }
\end{figure}
\begin{figure}
\caption{Experimental data (squares) and the analytic
         approximation (Eq.\ \protect\ref{analytic}, solid line)
         for the Raman intensity of CuGeO$_3$
         in the spin-Peierls phase at $T=5$ K. The phonon
         lines have been subtracted from the experimental data.
\label{T5}
             }
\end{figure}
\begin{figure}
\caption{Temperature dependence of the relative 
         intensities of three Peierls-active
         Raman lines. The $30$ cm$^{-1}$ line (squares),
         the $226$ cm$^{-1}$ line (triangles) and the
         $369$ cm$^{-1}$ line (circles, a folded phonon line) 
         The intensities are normalized
	 to their respective values at $T=6$ K. 
	 The intensities at $T=200$ K are
         subtracted for reference.
\label{rel}
             }
\end{figure}

\begin{references}
\bibitem{note_2} The Gibbs oscillations are suppressed by the
                 replacement $a_l\rightarrow a_l g(z_l)$
                 with $g(z_l) = [\sin(\pi z_l)/(\pi z_l)]^3$ and
                 $z_l = l/(N_p+1)$ \protect\cite{Silver}.
\bibitem{note_xy} A simple calculation for the dimerized $XY$-chain
                  shows that a vanishing matrix element in 
		  the Raman operator 
                  Eq.\ (\protect\ref{H_R}) removes this singularity.
\end{references}
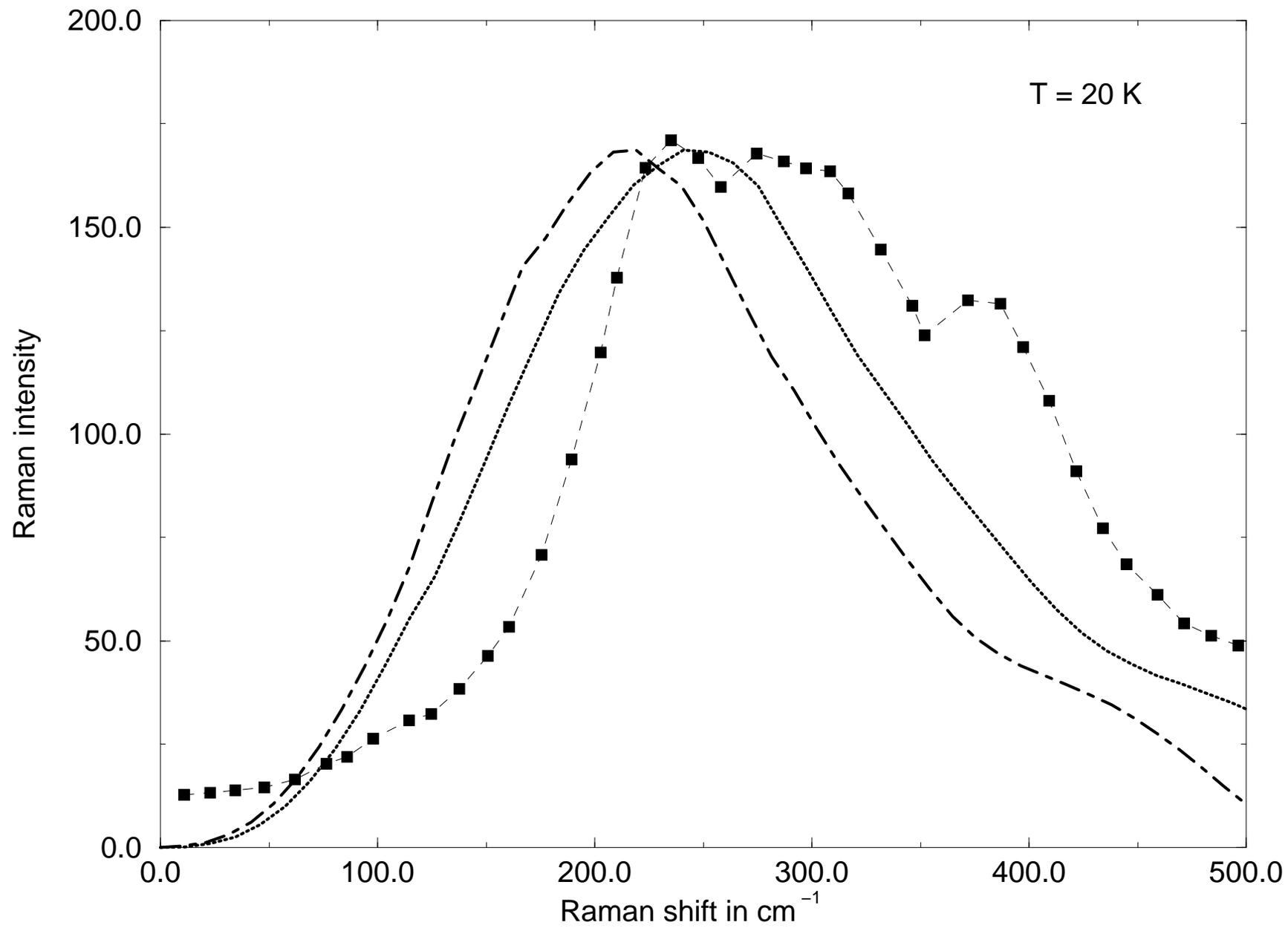
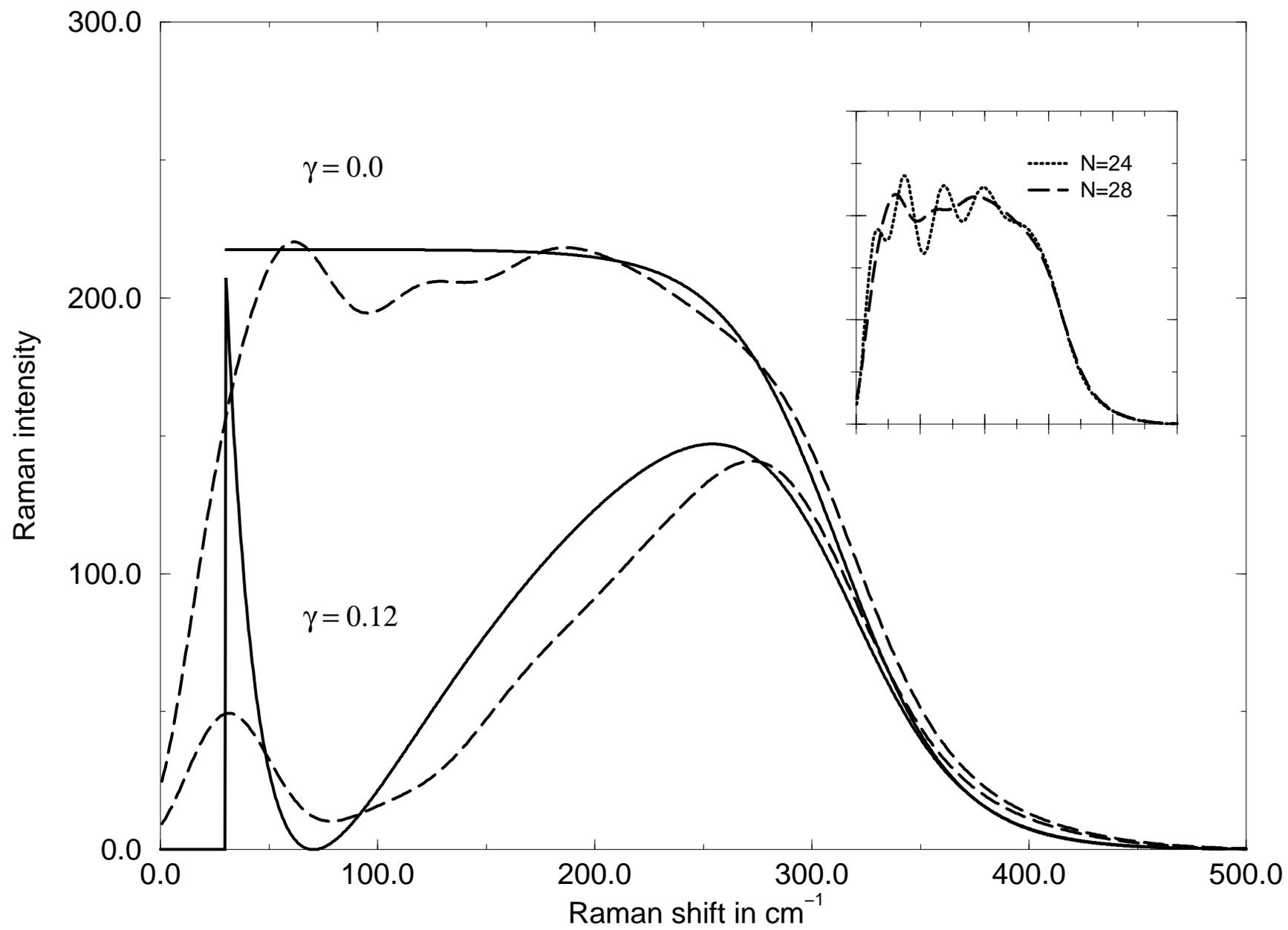
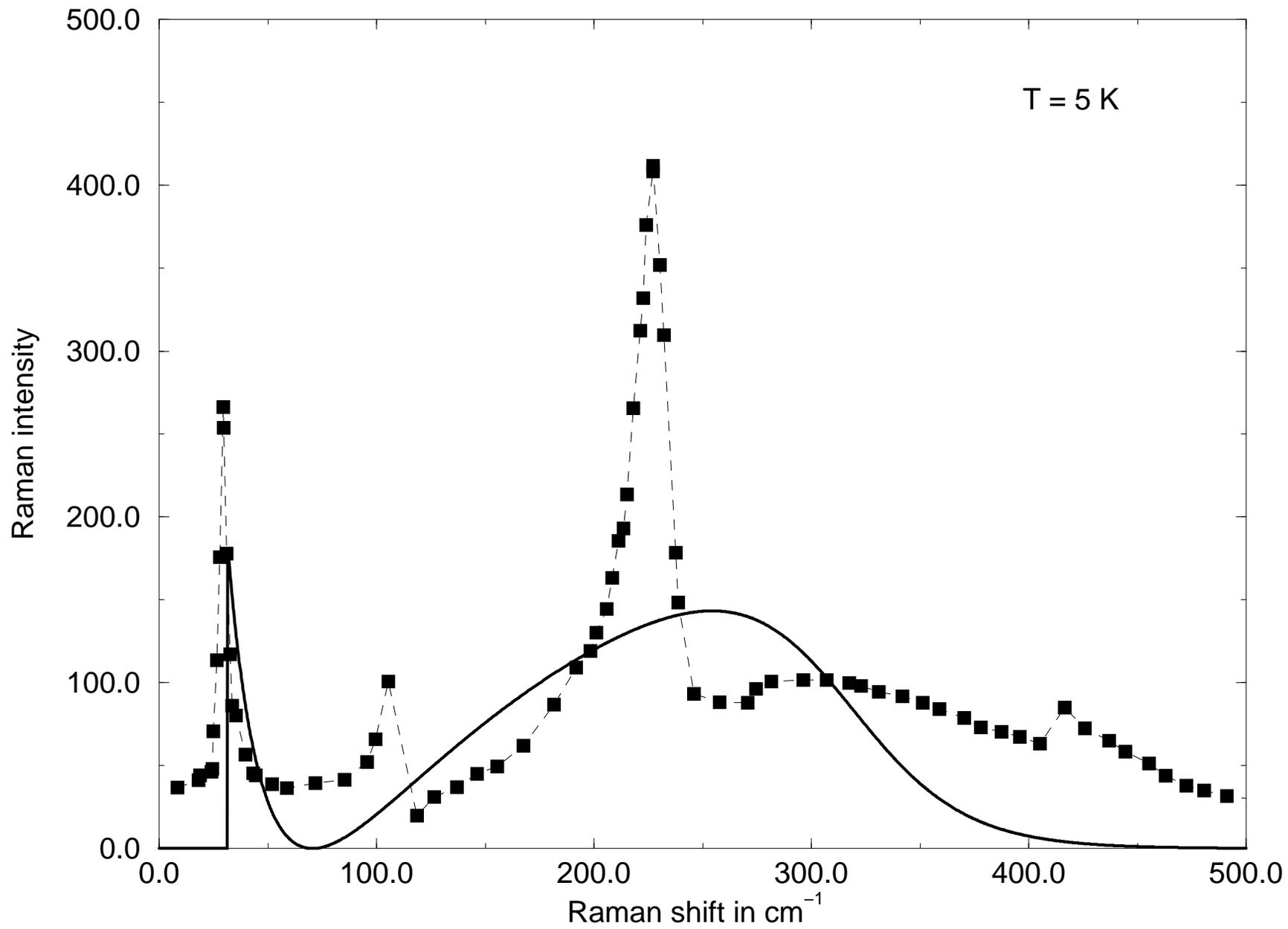
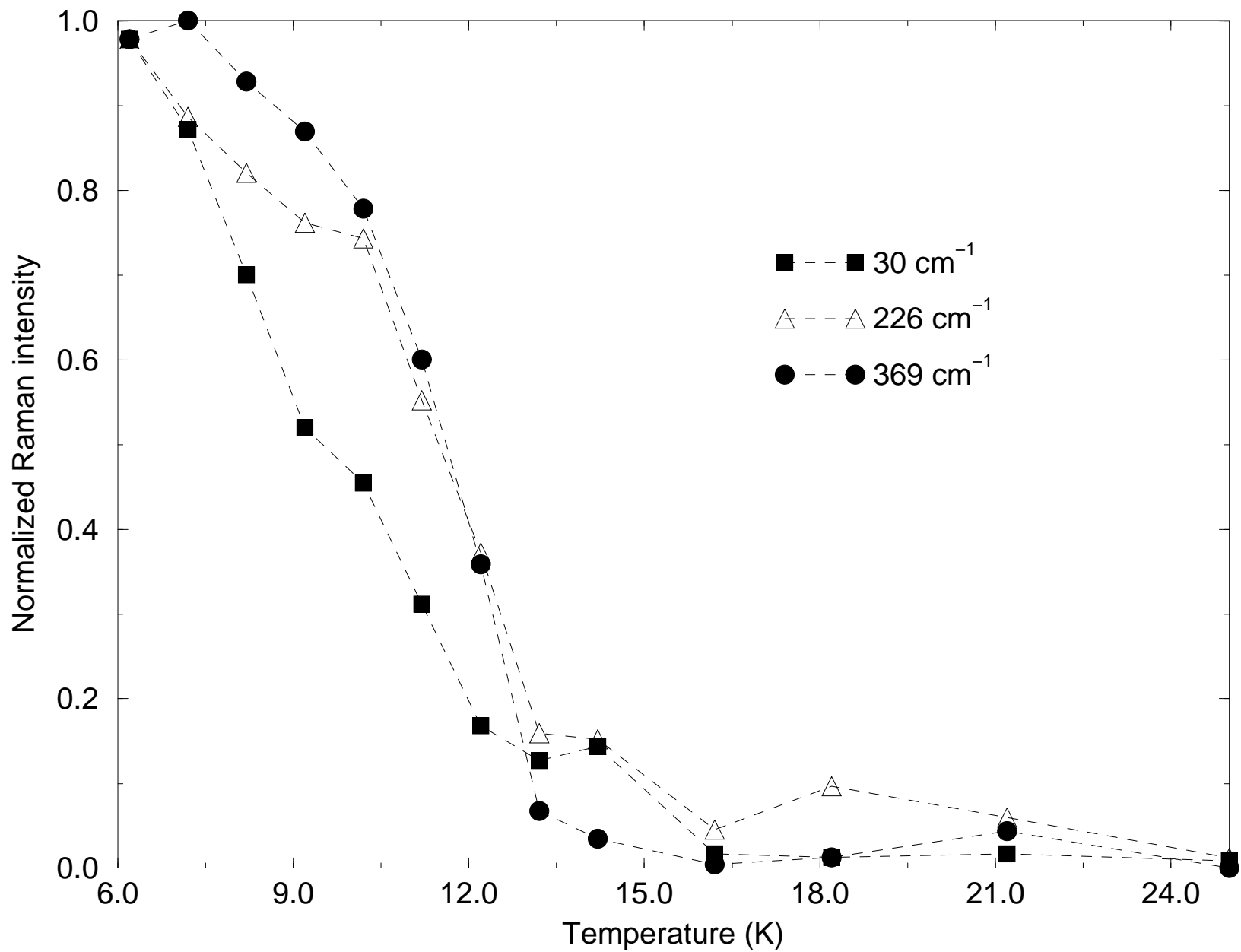
\end{document}